\documentclass[floatfix,showpacs,pra,aps,twocolumn]{revtex4}
\usepackage{bm}
\usepackage{amsmath}
\usepackage{amssymb}
\usepackage{latexsym} 
\usepackage{amsfonts} 
\usepackage{epsfig}
\usepackage{psfrag} 
\usepackage[latin1]{inputenc}
\usepackage[english]{babel}
\usepackage{amsmath}
\usepackage{amsfonts}
\usepackage{amssymb}
\usepackage{mathrsfs}
\usepackage{amsthm}

\newcommand{\ket}[1]{\left|#1\right>} 
\newcommand{\bra}[1]{\left<#1\right|} 
 
\newcommand{\bracket}[2]
 {\langle#1|#2\rangle} 
\newcommand{\nn}{\nonumber\\} 
 
\newcommand{\f}[1]{\mbox{\boldmath$#1$}}

\newcommand{\bea}{\begin{eqnarray}}
\newcommand{\ea}{\end{eqnarray}}
\newcommand{\eea}{\end{eqnarray}}
\newcommand{\ord}{{\cal O}}

\newcommand{\abs}[1]{\left|#1\right|} 
\newcommand{\pdiff}[2]{\frac{\partial#1}{\partial#2}}

\begin{document}

\title{Decoherence in a dynamical quantum phase transition}  


\author{Sarah Mostame$^{1,2}$, Gernot Schaller$^{3}$, and Ralf Sch\"utzhold$^{2,4 \, *}$}

\affiliation{$^1$Max-Planck Institut f\"ur Physik Komplexer Systeme, 
D-01187 Dresden, Germany
\\
$^2$Institut f\"ur Theoretische Physik, 
Technische Universit\"at Dresden, 
D-01062 Dresden, Germany
\\
$^3$Institut f\"ur Theoretische Physik,
Technische Universit\"at Berlin, 
D-10623 Berlin, Germany
\\
$^4$Fachbereich Physik, Universit\"at Duisburg-Essen, 
D-47048 Duisburg, Germany }

\begin{abstract} 
Motivated by the similarity between adiabatic quantum 
algorithms and quantum phase transitions, we study the impact of 
decoherence on the sweep through a second-order quantum phase transition
for the prototypical example of the Ising chain in a transverse field
and compare it to the adiabatic version of Grovers search algorithm, which displays a first order quantum phase transition.
%
For site-independent and site-dependent 
coupling strengths as well as different operator couplings, the results show that
(in contrast to first-order transitions) 
the impact of decoherence caused by a weak coupling
to a rather general environment increases with system size 
(i.e., number of spins/qubits).
This might limit the scalability of
the corresponding adiabatic quantum algorithm. 
\end{abstract} 


\maketitle
 
\section{Introduction}

\subsection{Quantum Phase Transitions}

%
In contrast to thermal phase transitions occurring when
the strength of the thermal fluctuations reaches
a certain
threshold, during recent years, a different class
of phase transitions has attracted the attention of physicists, 
namely transitions taking place at zero temperature \cite{sachdev}. 
%
An analytic 
non-thermal control parameter such as pressure, magnetic field, or
chemical composition is varied to access the transition point. 
%
Despite the analytic form of the order parameter, 
the ground state of a system changes non-analytically.
There, order is changed solely by quantum fluctuations, 
hence the name quantum phase transition (QPT).
Let us consider a quantum system (at zero temperature) described
by the Hamiltonian $H$ depending on some external
parameter $g$.
At a certain critical value of this parameter $g_c$,
the system is supposed to undergo a phase transition, i.e.,
the ground state $\ket{\Psi_{<} (g)}$ of $H(g)$ for $g < g_c$
is 
strongly
different from the ground state 
$\ket{\Psi_> (g)}$ of $H(g)$ for $g>g_c$. 
For example, $\ket{\Psi_< (g)}$ and $\ket{\Psi_> (g)}$
could have different global/topological properties 
(such as magnetization) in the 
thermodynamic limit.
%
\begin{figure}
\includegraphics[height=4.3cm,clip=true]{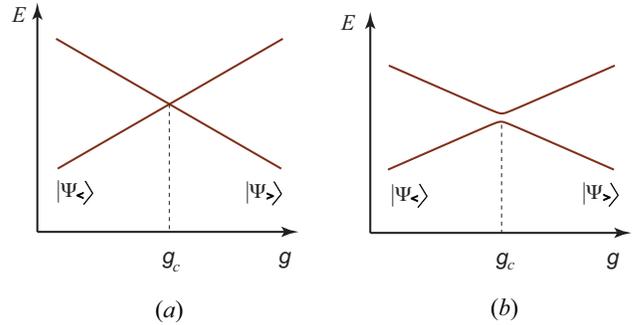}
\caption
{\label{phase-transition}
(Color Online)
Sketch of the lowest eigenvalues of a Hamiltonian $H(g)$ as
a function of some external parameter $g$
for a first order QPT.
At the critical point $g=g_c$, the ground state changes from 
$\ket{\Psi_< (g)}$ to $\ket{\Psi_> (g)}$.
$(a)$ A level-crossing. $(b)$ An avoided level-crossing.
}
\end{figure}
%
Therefore, a quantum phase transition  
can also be defined as a 
non-analyticity of the ground
state properties of the system as a function of the control parameter.
If this singularity arises from a simple level crossing in the 
ground state, see Fig.~(\ref{phase-transition})-$(a)$, 
then we have a first-order quantum phase transition.
The situation is different for continuous transitions, 
where a higher-order discontinuity in the ground state energy occurs.
Typically, for any finite-size system a 
transition will be 
rounded into a crossover, this is nothing but an avoided 
level-crossing in the ground state, see 
\mbox{Fig.~(\ref{phase-transition})-$(b)$}.
Continuous transitions can usually be characterized by an order 
parameter
%
which is a quantity that is zero in one phase (the disordered) 
and non-zero and possibly non-unique in the other (the ordered) phase. 
If the critical point is approached, the spatial correlations of 
the order parameter fluctuations become long-ranged. 
Close to the critical point the correlation length $\Upsilon$ 
diverges as
%
$\Upsilon^{-1} \propto \Lambda |g-g_c|^{\mathfrak{d}} \, ,$
%
where $\mathfrak{d}$ is a critical exponent and $\Lambda$ is an inverse
length scale of 
the order of 
the inverse lattice spacing.
Let $\Delta$ denote the 
smallest
energy excitation gap above the ground state.
In most cases, it has been found \cite{sachdev} that 
as $g$ approaches $g_c$, $\Delta$ vanishes as 
%
$\Delta \propto \Upsilon^{-z}
\propto \Lambda^{z} |g-g_c|^{\mathfrak{d} z}
\, ,$
where $z$ is the dynamic critical exponent.
This poses a scalability problem for adiabatic ground state preparation schemes (see below), as
these require a nonvanishing energy gap $\Delta$.

%
\subsection{Adiabatic Quantum Computation}
%

Unfortunately, the actual realization of usual sequential quantum
algorithms (where a sequence of quantum gates is applied to some
initial quantum state, see, e.g., \cite{nielsen}) goes along with
the problem that errors accumulate over many operations and the
resulting decoherence tends to destroy the fragile quantum features
needed for the computation.
Therefore, adiabatic quantum algorithms have been suggested
\cite{farhi-2000}, where the solution to a problem is encoded in the
(unknown) ground state of a (known) Hamiltonian.
Since there is evidence that, in adiabatic quantum computing 
the ground state is more robust against
decoherence -- the ground state cannot
decay and phase errors do not play any role, i.e., errors
can only result from excitations 
\cite{childs_robust,sarandy_open}-- this scheme 
offers fundamental advantages compared to sequential quantum algorithms
-- a sufficiently cold reservoir provided.
Suppose we have to solve a problem that may be 
reformulated as preparing a quantum system in the 
ground state of a Hamiltonian $H_{\rm f}$ . 
The adiabatic theorem \cite{messiah} then provides a
straightforward method to solve this problem:
Prepare the quantum system in the (known and easy-to-prepare) 
ground state of another Hamiltonian $H_0$.
Apply $H_0$ on the system and slowly 
modify it to $H_{\rm f}$.
The adiabatic theorem ensures for a non-vanishing time-dependent energy gap 
that if this has been done 
slowly enough, the system will end up in a state close to the ground state 
of $H_{\rm f}$.
Therefore, a measurement of the final state will 
yield a solution of the problem with high probability. 

Furthermore, adiabatic quantum
algorithms display a remarkable similarity with sweeps
through quantum phase transitions \cite{latorre,gernot}.
For all interesting systems discussed later in this article, 
adiabatic quantum computation inherently brings the
quantum system near to a point which is similar to 
the critical point in a quantum phase transition.
%
As an example 
for the deformation of the Hamiltonian,
one can consider the linear interpolation 
path between these two Hamiltonians
\bea
\label{adiabatic-hamiltonian}
H(g)=[1-g(t)]H_0+g(t)H_{\rm f}
\, ,
\ea
with $g(0)=0$ and $g(T)=1$, where
$T$ is the total evolution time or the {\em run-time}
of the algorithm.
%
%
We prepare the ground state of $H_0$ at time $t=0$,
and then the state evolves from $t=0$ to $T$ according 
to the Schr\"odinger equation.
At time $T$, we measure the state. 
According to the adiabatic
theorem, if there is a nonzero gap between the
ground state and the first excited state of $H(g)$
for all $g\in[0,1]$ then the success 
probability
of the algorithm approaches 1 in the limit $T\to\infty$.
How large $T$ should actually be is roughly given by \cite{sarandy2004} (for a more detailed discussion
see, e.g. \cite{schaller2006b,jansen2007a})
\bea
\label{adiabatic-condition0}
T\gg\frac{\max_{g\in[0,1]} 
\left|\bra{1,g}
\frac{dH(g)}{dg}
\ket{0,g}
\right|}
{\min_{g\in[0,1]}
\left[E_1(g)-E_0(g)\right]^2}\,,
\ea
where $E_0(g)$ is the lowest eigenvalue of 
$H(g)$, $E_1(g)$ is the second-lowest eigenvalue, 
and $\ket{0,g}$ and $\ket{1,g}$ are the corresponding
eigenstates, respectively.
Somewhere on the way from the simple initial 
configuration $H_0$ to the unknown solution of some problem 
encoded in $H_{\rm f}$, there is typically a critical 
point which bears strong similarities 
to a quantum phase transition.
%
At this critical point 
the fundamental gap 
(which is sufficiently large initially and finally) becomes 
very small, see, e.g., Fig.~(\ref{First-Grover}). 
Near the position of the minimum gap, the ground state 
will change more drastically
than in other time intervals of the interpolation.
In the continuum limit, one would generally expect that 
the minimum value of the fundamental gap in adiabatic 
computation will vanish
identically and that the ground state will change 
non-analytically at the critical point.
This is similar to what happens in quantum phase transition
when $g$ approaches $g_c$.
%
Based on this similarity, it seems \cite{gernot} that adiabatic
quantum algorithms corresponding to second-order quantum phase
transitions should be advantageous compared to isolated avoided level
crossings (which are analogous to first-order transitions).
A brief review of this idea comes in the following section.

\section{Examples}
%

\subsection{First-Order Transition -- Grovers Algorithm}
\begin{figure}
\includegraphics[height=5.5cm,clip=true]{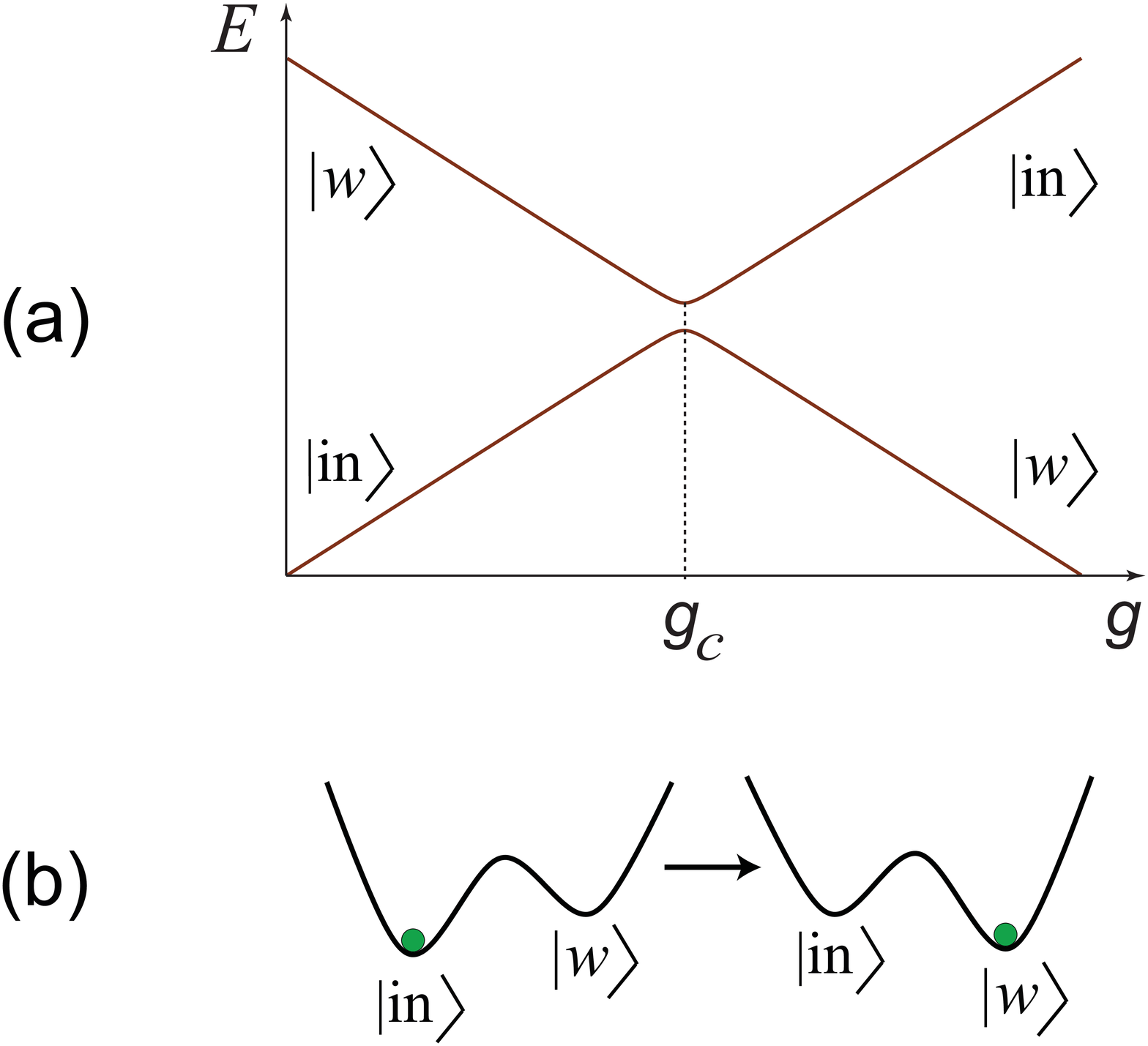}
\caption
{\label{First-Grover}
(Color Online)
Sketch of the two lowest energy eigenvalues of the Grover
Hamiltonian (a).
In the continuum limit, this corresponds to
the time evolution of
the energy landscape for a first-order transition (b). 
The green dot in the energy landscape denotes the ground state.
}
\end{figure}
%

\begin{figure}
\includegraphics[height=6cm,clip=true]{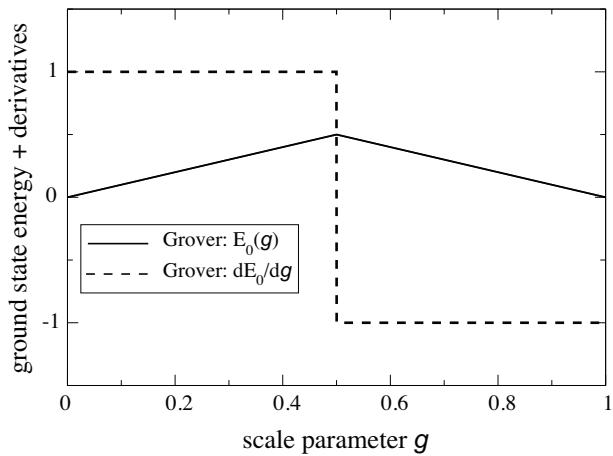}
\caption
{\label{grover-continuum}
The ground-state energy of the Grover Hamiltonian 
and its first-order derivative.
The discontinuity in the first-order derivative of the ground state
suggests the first-order quantum phase transition in 
adiabatic quantum search algorithm.
}
\end{figure}

An adiabatic version of Grovers algorithm \cite{roland} is defined by the Hamiltonian
\bea
\label{grover-hamiltoninan-3}
H(g)=(1-g)H_0+g H_{\rm{f}}
\, ,
\ea
where the initial Hamiltonian is given by 
\mbox{$H_0=\f{1}-\ket{{\rm in}}\bra{{\rm in}}$}
with the initial superposition state 
\mbox{$\ket{{\rm in}}=\sum_{x=0}^{D-1}\ket{x}/\sqrt{D}$}
and $D\equiv 2^N$ denotes the dimension of the Hilbert space for $N$ qubits.
The final Hamiltonian reads
\mbox{$H_{\rm f}=\f{1}-\ket{w}\bra{w}$}, 
where $\ket{w}$ denotes the marked state.
In this case, the commutator is very small 
$[H_0,H_{\rm f}]=
(\ket{{\rm in}}\bra{w}-\ket{w}\bra{{\rm in}})/\sqrt{D}$
and one can nearly diagonalize both Hamiltonians 
simultaneously and the $g$-dependent spectrum
will consist of nearly straight lines -- except near $g_c=1/2$,
where we have an avoided level-crossing, see Fig.~(\ref{First-Grover}).
In the continuum limit of $N\to\infty$, 
this corresponds to a first-order 
quantum phase transition from 
$\ket{\rm{in}}=\ket{\rightarrow \cdots \rightarrow}$
to
$\ket{w}=\ket{\uparrow \downarrow \cdots \uparrow \downarrow \downarrow}$,
for example, 
at the critical point $g_c=1/2$.
Such a first-order transition is characterized by an abrupt change of
the ground state -- $\ket{\rm{in}}$ for \mbox{$g<g_c$} 
and $\ket{w}$ 
for \mbox{$g>g_c$} 
-- resulting in a discontinuity of a corresponding order parameter,
see  Fig.~(\ref{grover-continuum}),
\bea
\label{first-derivative}
\bra{\psi_0(g)}\frac{dH}{dg}\ket{\psi_0(g)}
=\frac{dE_0}{dg}
\, .
\ea
In contrast to the conventional order parameters, for linear interpolations
in Eqn.~(\ref{grover-hamiltoninan-3}) 
the operator $dH/dg$ treats both phases symmetrically. Since
$[H_I, H_F] \neq 0$ (for nontrivial systems), $dH/dg$ is off-diagonal in
either phase and thereby plays an equivalent role such as e.g.
magnetization.
%
%
%

First-order quantum phase transitions are typically associated 
with an energy  landscape pictured in Fig.~(\ref{First-Grover}), 
where the two competing ground states are separated by an energy 
barrier throughout the interpolation.
In order to stay in the ground state, the system has to tunnel 
through the barrier between the initial ground state $\ket{\rm{in}}$ 
and the final ground state $\ket{w}$
during the quantum phase transition.
The natural increase of the strength of the barrier with the system size 
$N$ yields to the tunneling time which scales exponentially with the 
system size. 
Specifically, the optimal run-time for the adiabatic search 
algorithm behaves as \mbox{$T=\ord(\sqrt{D})=\ord\left(2^{N/2}\right)$} \cite{roland}.
The observation that this first order QPT is associated with an 
exponentially small energy gap right at the avoided crossing 
can be generalized \cite{schuetzhold2008a} to local Hamiltonians:
The two-dimensional subspace of the avoided crossing is spanned by
the eigenstates $\ket{w_<(g)}$ and $\ket{w_>(g)}$ that become degenerate
at $g=g_c$. 
Due to their macroscopic distinguishability, the overlap between these states
is exponentially small, which for local Hamiltonians also transfers to the
matrix element 
\mbox{$\bra{w_<(g_c)} H(g_c) \ket{w_>(g_c)} = \ord\left(\exp\{{-D}\}\right)$}.
Consequently, one can conclude from the eigenvalues of $H(g)$ in this two-dimensional
subspace
that also the minimum energy gap will become exponentially small in this case.

Therefore, the abrupt change of the ground state and the energy barrier
between the initial and final ground states suggest that the 
first-order transitions are not the best choice for the 
realization of adiabatic quantum algorithms \cite{gernot}. 
Thus, it would be relevant to study higher-order quantum phase 
transitions for this purpose.
%

\subsection{Second-Order Transition -- Ising Model}
%

The one-dimensional quantum Ising model is one of the two paradigmatic
examples \cite{sachdev} for second-order quantum phase transition (the other
is Bose-Hubbard model).
Of these two, only the former model is exactly solvable 
\cite{bunder,sachdev} -- the Ising model in a transverse field is a special case
of the XY model (which can also be diagonalized completely).
This  model has been employed in the study of quantum 
phase transitions and percolation theory \cite{sachdev}, spin glasses
\cite{sachdev,fischer}, as well as quantum annealing 
\cite{chakrabarti,santoro,kadowaki} etc. 
Although its Hamiltonian is quite simple, 
the Ising model is rich enough to display most of the basic phenomena
near quantum critical points. 
Furthermore, the transverse Ising model can also be used to study the
order-disorder transitions at zero temperature driven by quantum
fluctuations \cite{sachdev,chakrabarti}.
Finally, two-dimensional generalizations of the Ising model can be
mapped onto certain adiabatic quantum algorithms 
(see, e.g., \cite{dwave}). 
However, due to the evanescent excitation energies, such a phase 
transition is rather vulnerable to decoherence, which must be taken 
into account \cite{deco}.
%
\begin{figure}
\includegraphics[height=5.5cm,clip=true]{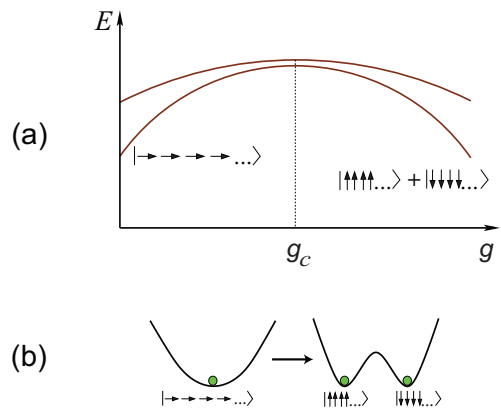}
\caption
{\label{Second-Ising}
(Color Online)
Sketch of the two lowest energy levels of the Ising
Hamiltonian (left) given in Eqn.~(\ref{Hamiltonian-Ising})
and the time evolution of the energy landscape for a 
second-order transition (right). 
A symmetry-breaking transition corresponds
to the deformation of the energy landscape.
The green dot in the energy landscape denotes the ground state.
}
\end{figure}
%

\begin{figure}
\includegraphics[height=6cm,clip=true]{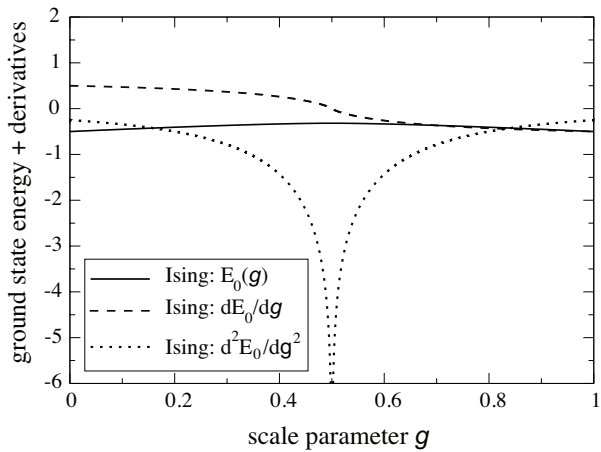}
\caption
{\label{ising-continuum}
The ground-state energy for the Ising model
and its first and second derivatives.
The discontinuity in the second-order derivative of the ground state 
suggests a second-order quantum phase transition.
}
\end{figure}

The one-dimensional transverse Ising chain of $N$ 
spins exhibits a time-dependent nearest-neighbor 
interaction $g(t)$ plus transverse field 
\mbox{$B(t)=1-g(t)$} 
\bea
\label{Hamiltonian-Ising}
{H}_{\rm{sys}}(t) 
= 
- \sum_{j=1}^N\left\{[1-g(t)]\,\sigma^x_j+
g(t)\,\sigma^z_j\sigma^z_{j+1}\right\} 
\,, 
\ea
where \mbox{${\f\sigma}_j=(\sigma^x_j,\sigma^y_j,\sigma^z_j)$} are the
spin-1/2 Pauli matrices acting on the $j$th qubit and periodic boundary
conditions \mbox{${\f\sigma}_{N+1}={\f\sigma}_1$} are imposed. 
This Hamiltonian is  invariant under a global 180-degree rotation around 
the $\sigma_j^x$-axes (bit flip) which transforms all qubits according to 
\mbox{$\sigma_j^z \to -\sigma_j^z$}.
%
%
Choosing \mbox{$g(0)=0$} and \mbox{$g(T)=1$} 
where $T$ is the evolution time,  
the quantum system evolves from the unique paramagnetic state 
\mbox{$\ket{\rm{in}}=\ket{\rightarrow\rightarrow\rightarrow\dots}$} 
through a second-order quantum phase transition
(see figure \ref{ising-continuum})
at \mbox{$g_{\rm cr}=1/2\, \, $} to 
the symmetrized combination in the two-fold degenerate ferromagnetic subspace, 
see also Fig.~\ref{Second-Ising}.
\bea\label{Esupercat}
\ket{w}=\frac{\ket{\uparrow\uparrow\uparrow\dots} 
+\ket{\downarrow\downarrow\downarrow\dots}}{\sqrt{2}}
\, .
\ea
At the critical point $g_{\rm cr}$ the excitation gap
vanishes (in the thermodynamic limit $N\to\infty$) and the
response time diverges. 
As a result, driving the system through its quantum critical point at
a finite sweep rate entails interesting non-equilibrium phenomena such 
as the creation of topological defects, i.e., kinks \cite{dziarmaga}. 
%
%

Since the initial ground state $\ket{\rm{in}}$ reflects the 
bit-flip invariance of the Hamiltonian~(\ref{Hamiltonian-Ising}) 
whereas the final ground state subspace 
$\ket{\uparrow\ldots\uparrow}$ and $\ket{\downarrow\ldots\downarrow}$ 
breaks this symmetry, we have a 
symmetry-breaking quantum phase transition.
Typically, such a symmetry-breaking change of the ground state
corresponds to  a second-order phase transition \cite{gernot}.
For such a transition, the ground state changes 
continuously and the energy barrier observed in first-order transitions is absent:
Initially, there is a unique ground state 
but at the critical point, this ground state splits up into 
two degenerate ground states which are the mirror
image of each other. 
Therefore, the ground state
does not change abruptly in this situation and the system
does not need to tunnel through a barrier in order
to stay in the ground state, 
see Figs.~(\ref{Second-Ising}b). 
Consequently, we expect that in this case,
a closed quantum system should find its way from the initial to the 
final ground state easier.
This expectation is confirmed in 
the following sections of this article: 
Since the minimum gap behaves as $\ord(1/N)$,
the optimal run-time in order to stay in the 
ground state scales polynomially for the Ising model.

\subsection{Mixed Case}

\begin{figure}
\includegraphics[height=2cm,clip=true]{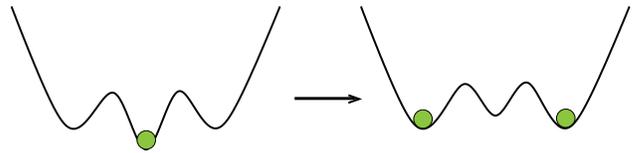}
\caption
{\label{mixed-first}
(Color Online)
Sketch of the time evolution of the energy landscape
for a symmetry-breaking quantum phase transition which
is of first order.
The green dot denotes the ground state.
}
\end{figure}

\begin{figure}
\includegraphics[height=6cm,clip=true]{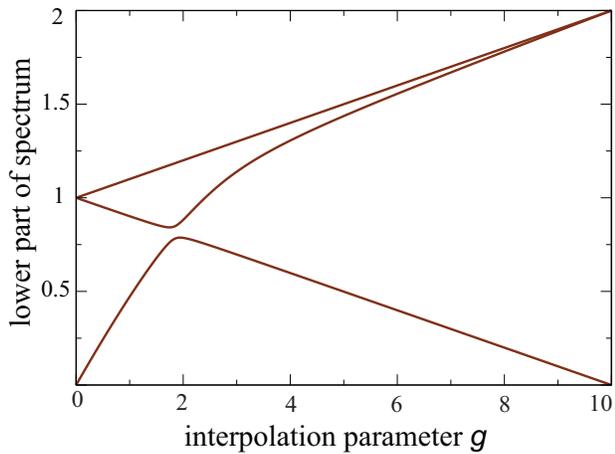}
\caption
{\label{mixed-energy}
(Color Online)
Sketch of the lowest eigenvalues of the 
Hamiltonian~(\ref{mixed-hamiltonian}).
One can clearly see that the spectrum displays an avoided-level 
crossing at the critical point -- thus corresponding to a 
first-order transition \cite{gernot-symmetry}.
}
\end{figure}

Looking at Fig.~(\ref{Second-Ising}), it seems that a symmetry-breaking 
quantum phase transition typically corresponds to a 
second-order phase transition, but there are counter-examples:
Consider more complicated energy landscape \cite{gernot-symmetry} in 
Fig.~(\ref{mixed-first}).
%
%
In spite of the 
mirror symmetry of the energy landscape,
there is a tunneling barrier throughout
the interpolation.
An analytic example for a symmetry-breaking first-order transition
\cite{gernot-symmetry} 
is given by a combination of the initial Hamiltonian from the Grover
problem with the final Hamiltonian of the Ising model
\bea
\label{mixed-hamiltonian}
H_0=\f{1}-\ket{{\rm in}}\bra{{\rm in}}
\quad
,
\quad
H_{\rm f} = \frac{1}{2}
\sum_{j=1}^N
\left(\f{1} -\sigma_j^z\sigma_{j+1}^z
\right)
\, ,
\ea
where $H_{\rm f}$ has been shifted and scaled in order to 
preserve the positive definiteness.
Even though this Hamiltonian possesses
the same bit-flip symmetry as the Ising model, 
its level structure displays an avoided-level 
crossing at the critical point, i.e., it corresponds 
to a first-order phase transition
with a jump between the initial and the final
ground state(s), see 
Fig.~(\ref{mixed-energy}).

It can also be shown analytically that the fundamental gap of the
combined Hamiltonian $H(g)=(1-g)H_0+g H_{\rm f}$ 
vanishes exponentially with the system size, 
i.e. the number of qubits, see, e.g., \cite{farhi0512159,zindaric-NP}.
%

\section{Decoherence in the Adiabatic Limit}

In all of the above examples, we have
seen that at the critical point, at least some energy levels become 
arbitrarily close and thus, the response times diverge 
(in the continuum limit).
Consequently, during the sweep through such a phase transition by
means of a time-dependent external parameter, small external
perturbations or internal fluctuations become strongly amplified -- 
leading to many interesting effects, 
see, e.g., 
\cite{group_phase0,group_phase1,vidal,zurek,damski,sengupta,fazio}.
One of them is the anomalously high susceptibility to decoherence 
(see also \cite{fubini}):
Due to the convergence of the energy levels at the critical point,
even low-energy modes of the environment may cause excitations and
thus perturb the system. 
Based on the similarity between the quantum adiabatic algorithms and 
critical phenomena, we have argued that adiabatic quantum algorithms 
corresponding to the higher-order quantum phase transitions should be 
advantageous in comparison to those of first order for closed quantum systems.
The present paper aims at generalizations to these findings when the
impact of decoherence is considered.

In order to study the impact of decoherence,
we consider an open system described by the total
Hamiltonian ${H(t)}$ which can be split up into that of the closed
system ${H}_{\rm{sys}}$ and the bath ${H}_{\rm{bath}}$ acting
on independent Hilbert spaces  
\mbox{${\mathcal H}_{\rm{sys}}\otimes{\mathcal H}_{\rm{bath}}={\mathcal H}$} 
\bea
\label{HTotal}
{H}(t) \, = \,  
{H}_{\rm{sys}} (t) + {H}_{\rm{bath}} + \lambda \, {H}_{\rm{int}}
\,, 
\ea
plus an interaction \mbox{$\lambda \, {H}_{\rm{int}}$} between the two,
which is supposed to be weak \mbox{$\lambda\ll1$} in the sense that it does
not perturb the state of the system drastically.  
Note that the change of the bath caused by the interaction 
need not be small.   
To describe the evolution of the combined quantum state 
\mbox{$\ket{\Phi(t)}\in{\mathcal H}$}, we expand 
it
using the instantaneous system energy eigenbasis 
\mbox{${H}_{\rm{sys}}(t)\ket{\psi_s(t)}=E_s(t)\ket{\psi_s(t)}$},
via
\bea
\label{QState}
\ket{\Phi(t)}=
\sum_s a_s(t)\ket{\psi_s(t)}\otimes\ket{\alpha_s(t)}\, ,
\ea
where $a_s$ are the corresponding amplitudes and 
\mbox{$\ket{\alpha_s}\in{\mathcal H}_{\rm{bath}}$} denote the associated  
(normalized but not necessary orthogonal) states of the reservoir.  
Insertion of this expansion into the Schr\"{o}dinger equation  
\mbox{$i\ket{\dot{\Phi}(t)}={H}(t)\ket{\Phi (t)}$} yields 
\mbox{($\hbar=1$)}
\bea 
\label{Diff}  
&&
\pdiff{}{t} 
\left(a_s e^{-i\varphi_s}\right) 
= 
e^{-i\varphi_s}
\sum_{r\ne s} a_r \, \times
\nn
&&
\hspace{-1.1cm}
\left(   
\frac{\bra{\psi_s }\dot{H}_{\rm{sys}}\ket{\psi_r}}{\Delta E_{sr}} 
\,\bracket{\alpha_s}{\alpha_r}  
-i
\bra{\alpha_s}\bra{\psi_s}\lambda{H}_{\rm{int}}\ket{\psi_r}\ket{\alpha_r}
\right)
\ea 	
with the energy gaps \mbox{$\Delta E_{sr}(t)=E_s(t)-E_r(t)$} of the
system and the total phase (including the Berry phase)
\bea
\label{Phase}
&&
\varphi_s(t) 
=
- \int_0 ^t dt' 
\Big\{ 
E_s (t') +
H_{\rm{bath}}^{ss} (t') +
\lambda H_{\rm{int}}^{ss} (t') 
\nn
&&
\hspace{0.85cm}
-i\bracket{\psi_s(t')}{\dot{\psi}_s(t')} 
-i\bracket{\alpha_s (t')}{\dot{\alpha}_s(t')}  
\Big\} 
\ea     
with the energy shift  
$H_{\rm{int}}^{sr}=\bra{\alpha_s}\bra{\psi_s}
{H}_{\rm{int}}\ket{\psi_r}\ket{\alpha_r}$
, 
$H_{\rm{bath}}^{sr}=
\bra{\alpha_s}{H}_{\rm{bath}}\ket{\alpha_r}$.
Evidently, there are two contributions for transitions in the Hilbert
space ${\mathcal H}_{\rm{sys}}$ of the system:
\begin{enumerate}
\item[-]
The first term on the right-hand side of Eqn.~(\ref{Diff}) describes
the transitions caused by a non-adiabatic evolution \cite{sarandy2004}.
Note, however, that the factor \mbox{$\bracket{\alpha_s}{\alpha_r}$} and the
additional phases in Eqn.~(\ref{Phase}) give rise to 
modifications in the adiabatic expansion. 
\item[-]
The last term in Eqn.~(\ref{Diff}) directly corresponds to transitions
caused by the interaction of the quantum system with its environment.
\end{enumerate}
Since we are mainly interested in the impact of the coupling to the
bath, we shall assume a perfectly adiabatic 
evolution of the system itself, 
i.e., without the coupling to the environment 
$\lambda=0$, the system would stay in its ground state. 
Thus, the only decoherence channel available is heating 
(i.e., excitations), the phase damping and decay channels, for
example, play no major role here. 
Considering the adiabatic condition
%
\bea
\label{adiabatic-cond} 
\bra{\psi_s}\dot{H}_{\rm{sys}}\ket{\psi_r}\ll\left(\Delta E_{sr}\right)^2
\, ,
\ea
the first term in Eqn.~(\ref{Diff}) is negligible and the
second one dominates. 
Starting in the system's ground state \mbox{$a_0(t=0)=1$}, 
which is relevant for adiabatic quantum computation, 
the excitations \mbox{$s>0$} caused by the weak interaction 
\mbox{$\lambda H_{\rm{int}}$} with the bath 
\bea
\label{excitation00}
\mathfrak{A}_s \equiv a_s(T) \exp\{-i\varphi_s(T)\}
\ea
can be calculated via response theory, i.e., the solution of Eqn.~(\ref{Diff}) 
is to first order in \mbox{$\lambda\ll1$} given by
\bea
\label{GAmplitude}
\mathfrak{A}_s 
\approx
- i 
\int_0 ^T dt \, \, 
e^{i \Delta \varphi_{0s}}  
\bra{\alpha_s}  
\bra{ \psi_s} 
\lambda H_{\rm{int}} 
\ket{\psi_0} 
\ket{\alpha_0} 
\,, 
\ea
where \mbox{$ \Delta \varphi_{rs} = \varphi_r (t) - \varphi_s (t)$}.
This is rather a general result.

In the following, after a brief review of the impact of decoherence
on the sweep through a first-order quantum phase transition
\cite{markus} in section \ref{Sdec_grover},
we study in section \ref{Sdec_ising} the impact of decoherence due to
a general reservoir for the quantum Ising chain in a transverse
field, which is considered a prototypical example   
for a second-order quantum phase transition.

\section{Decoherence in the Adiabatic Grover Search}\label{Sdec_grover}

Let us consider the Grover model~(\ref{grover-hamiltoninan-3}) -- weakly coupled to a bath, 
where we can assume the following expansion of the 
interaction Hamiltonian \cite{markus}
\bea
\label{grover-interaction}
\lambda H_{\rm{int}} = 
\lambda\sum_{j=1}^N \f{\sigma}_j \cdot \f{A}_j
+\lambda^2\sum_{\ell,j=1}^N 
\f{\sigma}_{\ell} \cdot \f{B}_{\ell j}
\cdot\f{\sigma}_j 
+\ord(\lambda^3)
\, ,
\nn
\ea
where $\lambda \ll 1$ and 
\mbox{$\f{\sigma}_j (t) = (\sigma_j^x(t),\sigma_j^y(t),\sigma_j^z(t))$}
is the vector of Pauli matrices in the interaction 
picture with the corresponding bath operators $\f{A}_j (t)$ and 
$\f{B}_{\ell j} (t)$, etc.
Recalling the adiabatic version of Grovers search algorithm in 
Eqn.~(\ref{grover-hamiltoninan-3}), at the beginning of the 
evolution the system has to 
be 
prepared in the ground state  
\mbox{$\ket{{\rm in}}=\sum_{x=0}^{D-1}\ket{x}/\sqrt{D}$}.
This also requires that the initial full density operator can be initialized
as a direct product
\bea
\label{initial-density-op}
\varrho(0)=\varrho_{\rm{sys}} (0) \otimes \varrho_{\rm{bath}} (0) 
\, , 
\ea
i.e., system and environment are not entangled at the beginning.
Since in the weak-coupling limit the
adiabaticity condition for the open system dynamics is still in
leading order the same as for the closed system, similar to 
the discussions in the former section, one can 
assume perfect adiabatic evolution
of the unperturbed system and hence only consider 
perturbations due to the interaction with the environment. 

The spectrum of the Grovers Hamiltonian~(\ref{grover-hamiltoninan-3})
consists of the ground state $\ket{\psi_0(g)}$ and the first excited state
$\ket{\psi_1(g)}$, which come very close 
$\left(\Delta E_{\rm{min}} =1/\sqrt{D}\right)$
at \mbox{$g_c=1/2$},  
whereas all other states $\ket{\psi_{k>1} (g)}$ 
are degenerate and well separated from the ground state by an energy gap of order
one.
Since the temperature and hence the energies
available in the environment 
must 
be much smaller than that gap of order one 
(in order to prepare the initial ground state), 
transitions from the
ground state to these states $\ket{\psi_{k>1}(g)}$ are exponentially 
suppressed.
Thus, the final probability of the transitions to the 
first excited state  \cite{markus}
\bea
\label{prob0}
\left|\mathfrak{A}_s \right|^2
&\approx&
\lambda^2\sum_{
\begin{subarray}{c}
\ell,j=1 \\
\mu,\nu =x,y,z
\end{subarray}}^N
\int_0 ^T dt_1
\int_0 ^T dt_2 \, 
\left<
A_{\ell}^{\mu} (t_1)
A_j^{\nu} (t_2)
\right>
\times
\nn
&& 
\times \, 
\bra{w^{\perp}}\sigma_{\ell}^{\mu} (t_1)\ket{w}
\bra{w}\sigma_j^{\nu} (t_2)\ket{w^{\perp}}
\, ,
\ea
provides a good measure for the success 
probability which corresponds to 
\mbox{$\left|\mathfrak{A}_s \right|^2 \ll 1$}.
It can be shown \cite{markus} that the contributions proportional to $\f{B}_{\ell j}$ do not
contribute to second order in $\lambda$.
In this equation, $\ket{w}$ denotes the marked state for Grovers problem,
see Eqn.~(\ref{grover-hamiltoninan-3}), 
and $\ket{w^\perp}$ is the state orthogonal to $\ket{w}$ in the
subspace spanned by $\ket{w}$ and $\ket{\rm in}$.
The expression~(\ref{prob0}) demonstrates that both system and reservoir properties
affect the excitation amplitude.
Of the system matrix elements 
$\bra{w^{\perp}}\sigma_j^{\mu} (t)\ket{w}$
only those with $\mu=x,z$ contribute (for large 
$D=2^N \ggg 1 $,
the $\mu=y$ term is suppressed by a factor
$1/\sqrt{D}$) 
\bea
\label{matrix-system}
\bra{w^{\perp}}\sigma_j^x (t)\ket{w}
\approx
- \, \frac{1-g(t)}{\sqrt{D}\Delta E(t)} 
\exp\left\{-i\int_0^t dt'\Delta E(t') \right\}
\, ,
\nn
\ea
for large $N$ and
\mbox{$\Delta E(t) =\sqrt{1-4g(t)[1-g(t)](1-1/D)}$}. 
It is the same for \mbox{$\bra{w^{\perp}}\sigma_j^z (t)\ket{w}$}
apart from an additional sign 
$(-1)^{w_j +1}$, where $w_j$ is the $j$-th bit of $w$, i.e., 
$\ket{w}$ is an eigenstate of the operators 
$\sigma_j^z$ with eigenvalues $(-1)^{w_j}$.

Assuming a stationary reservoir 
$\left[H_{\rm{bath}},\varrho_{\rm{bath}}  \right]=0$ 
(which does not necessarily imply a bath in thermal equilibrium)
allows for a Fourier decomposition of the bath correlation function
\bea
\label{grover-fourier}
\left<
A_{\ell}^{\mu} (t_1)
A_j^{\nu} (t_2)
\right>
=\int_{-\infty}^{+\infty} d\omega \, e^{-i\omega (t_1-t_2)}
f_{\ell j}^{\mu \nu} (\omega)
\, , 
\ea
where $f_{\ell j}^{\mu \nu} (\omega)$ depends on the spectral 
distribution of the bath modes and the temperature, etc.

For example, for a bosonic bath in thermal equilibrium and coupling operators 
\mbox{$A_j^\mu \propto \sum_k \left[h_k a_k + h_k^* a_k^\dagger\right]$} 
(as e.g., used in the spin-boson model)
we would for an inverse bath temperature $\beta$ obtain a Fourier decomposition such as \cite{schaller2008a}
%
\bea
\label{Efourierexample}
f_{\ell j}^{\mu\nu}(\omega) \propto \frac{J(\abs{\omega})}{\abs{1-e^{-\beta \omega}}}
= \, J(\abs{\omega}) \left[ \frac{1}{e^{\beta \abs{\omega}} -1} + \Theta (\omega) \right] \,,
\nn
\ea
where $ \Theta (\omega)$ is the step function being 1 for $\omega>0$ and
0 for $\omega<0$.
$J(\omega)$ denotes the spectral density, which is often parameterized as \cite{brandes2005a}
%
\bea
\label{Especdens}
J(\omega) = 
2 \vartheta \omega_{\rm ph}^{1-\epsilon} \omega^\epsilon e^{-\omega/\omega_{\rm c}}\,,
\eea
where $0 \le \epsilon < 1$ corresponds to the sub-ohmic, 
$\epsilon=1$ to the ohmic, and $\epsilon>1$ to the super-ohmic case.

Insertion of~(\ref{matrix-system}) and~(\ref{grover-fourier}) into Eqn.~(\ref{prob0}) yields
\bea
\label{prob1}
\hspace{1cm}
\left|\mathfrak{A}_s \right|^2
&\approx&
\lambda^2
\int d\omega \sum_{\ell , j=1}^N
f_{\ell j}^{xx} (\omega)\, \times
\nn
&&
\hspace{-3cm}
\times \,
\left|
\int_0^T dt \, \, 
\frac{1-g(t)}{\sqrt{D}\Delta E(t)}
\exp\left\{i\omega t +i\int_0^t dt'\Delta E(t') \right\}
\right|^2
\ea
plus similar terms including $f_{\ell j}^{xz}$, 
$f_{\ell j}^{zx}$ and $f_{\ell j}^{zz}$ with the 
associated signs $(-1)$ and $(-1)^{w_j}$ for $x$
and $z$, respectively \cite{markus}.
In order to evaluate the time integrations, it is useful to
distinguish different domains of $\omega$:
\begin{enumerate}
\item[-] 
For large frequencies $|\omega|\gg \Delta E_{\rm{min}}$, 
the time integral can be calculated via the
saddle-point approximation.
The saddle points $t_{\omega}^*$  are given by a vanishing
derivative of the exponent
\bea
\label{saddle-point-grover}
\omega +\Delta E(t_{\omega}^*)= 0
\, ,
\ea
which corresponds to energy conservation.
Hence large positive frequencies
$\omega\gg \Delta E_{\rm{min}}$
do not contribute at all which is quite natural (this 
corresponds to the transfer of a large energy from the
system to the reservoir).
\item[-]
The saddle-point approximation cannot be applied for small
frequencies $\omega=\ord(\Delta E_{\rm{min}})$
and energy conservation is also not well-defined.
In this case, one might estimate an upper bound for the time 
integral by omitting all phases.
\end{enumerate}
These, altogether yield 
\bea
\left|\mathfrak{A}_s \right|^2
&\approx&
\lambda^2 D
\int _{-\Delta E_{\rm{min}}}^{+\Delta E_{\rm{min}}}
d\omega
\, f(\omega)
\nn
&&
+\, \frac{\pi\lambda^2}{2D}
\int _{\Delta E_{\rm{min}}}^1
d\omega
\, \frac{f(-\omega)}
{\omega^2\dot{g}(t_{\omega}^*)}
\, ,
\ea
where $f(\omega)$ is the appropriate sum of the $f_{\ell j}^{xx}$,
$f_{\ell j}^{xz}$, $f_{\ell j}^{zx}$ and $f_{\ell j}^{zz}$
contributions.
The second term of the above equation depends on the interpolation
function $g(t)$. 
Considering three scenarios \cite{schaller2006b}
%
\bea
{\rm{(a)} } \, \, \ddot{g}=0
\qquad
{\rm{(b)} } \, \, \dot{g}\propto\Delta E
\qquad
{\rm{(c)} } \, \, \dot{g}\propto\Delta E^2
\nonumber
\ea
%
the second integrand scales as
%
\bea
 {\rm{(a)} } \, \,  \frac{D \, f(-\omega)}{\omega^2}
\qquad
{\rm{(b)} } \, \, \frac{ \sqrt{D}\,f(-\omega)}{\omega^3}
\qquad
{\rm{(c)} } \, \, \frac{f(-\omega)}{\omega^4} \, ,
\nonumber
\ea
%
respectively.
In all of these cases, the bath modes with large frequencies
$|\omega|\gg \Delta E_{\rm{min}}$ do not cause problems in the
large-$N$($D$) limit, since
the spectral function $f(-\omega)$ is supposed to decrease for 
large $|\omega|$ as the bath does not contain excitations with 
large energies -- the environment is cold enough, 
compare also Eqn.~(\ref{Efourierexample}).
Therefore, the low-energy modes of the reservoir
$\omega=\ord(\Delta E_{\rm{min}})$ give the potentially 
dangerous contributions.
Independent of the dynamics $g(t)$ both the first integral
and the lower limit of the second integral yield
the same order of magnitude \cite{markus}
\bea
\label{grover-last-estimate}
\left|\mathfrak{A}_s \right|^2
\approx
\lambda^2\,  
\frac{f\left[\ord(\Delta E_{\rm{min}})\right]}
{\Delta E_{\rm{min}}}
\, .
\ea
Since $\Delta E_{\rm{min}}$ decreases as 
$1/\sqrt{D}$ in the large-$D$
limit, the spectral function $f(\omega)$ must vanish in the infrared limit 
as $\omega$ or even faster in order to keep the error $\left|\mathfrak{A}_s \right|^2$
under control.
Thus, one can conclude that
the spectral function $f(\omega)$ of the bath provides a 
criterion to favor or disfavor certain physical implementations. 
If $f(\omega)$ vanishes in the infrared limit faster than $\omega$ 
-- compare also Eqn.~(\ref{Especdens}) --
the computational error does not grow with increasing system size
-- {\em the quantum computer is scalable}.
%
This result has already been derived in \cite{markus}  with a slightly different formalism.

\section{Results: Decoherence in the Transverse Ising Chain}\label{Sdec_ising}

%
As we shall see below, the situation may be very different
for second-order transitions compared to the first-order 
transitions.
These investigations are particularly relevant in view of the 
announcement (see, e.g., the discussion in \cite{dwave}) regarding the construction of
an adiabatic quantum computer with 16 qubits in the form of a
two-dimensional Ising model.  

First of all, we briefly review the main steps \cite{sachdev} of the analytic diagonalization of 
${H}_{\rm{sys}}$, where we switch temporarily to the Heisenberg
picture for convenience:
The set of $N$ qubits in Eqn.~(\ref{Hamiltonian-Ising}) can be mapped to a system
of $N$ spinless fermions $c_j$ via the {\em Jordan-Wigner} transformation
\cite{jordan} given by  
\bea 
\label{Jordan}
&& 
\sigma_j ^x (t) 
= 
1 - 2 c_j ^{\dag} (t) \,  c_j (t) \,, 
\nn
&& 
\sigma_j ^z (t) 
= 
- \prod_{\ell < j} \, 
\left[1 - 2 c_{\ell} ^{\dag} (t) \, c_{\ell} (t) \right]  
\left[ c_j (t) + c_j ^{\dag} (t) \right]
\ea 
with 
\mbox{$ \bm\sigma_j (t) $} 
indicating the Pauli operators in Heisenberg picture 
\mbox{$\bm\sigma_j (t) = 
\mathcal{U}^{\dagger}_{\rm{sys}} (t) \, 
\bm\sigma_j \,  
\mathcal{U}_{\rm{sys}} (t)$}, 
where \mbox{$ \mathcal{U}_{\rm{sys}} (t) $}
is the unitary time evolution operator of the system. 
It is easy to verify that the fermionic operators 
anti-commutation relations satisfy
\bea
\label{jordan-wigner2}
\left\{c_{\ell}, c_j^{\dag} \right\}=\delta_{\ell j}
\quad
,
\quad
\left\{c_{\ell}, c_j \right\}
=
\left\{c_{\ell} ^{\dag}, c_j^{\dag} \right\}
=0
\, .
\ea
Insertion of Eqn.~(\ref{Jordan}) into the system Hamiltonian in 
Eqn.~(\ref{Hamiltonian-Ising}) yields
in the subspace of an even particle number 
%
\bea
\label{HJordan}
&&
H_{\rm{sys}} (t) 
= \,
- \sum_{j=1}^N  \Big\{ \left[1 - g(t) \right] 
\left( 1 - 2 c_j ^{\dag} \, c_j \right)\,\\
&& 
+ \,   g(t)  \left(  c_{j+1} \, c_j
+ c_{j+1} ^{\dag}  \,  c_j  
+ c_j ^{\dag}  \, c_{j+1}  
+ c_j ^{\dag} \, c_{j+1} ^{\dag}  \right)\Big\}
\,,\nonumber
\ea
where the time-dependency of the $c_j$ has been dropped for brevity. 
This fermionic Hamiltonian has terms that violate the fermion 
conservation number,
$ c_{j+1} \, c_j$ and $c_j ^{\dag} \, c_{j+1} ^{\dag} $. 
This bilinear form can now be diagonalized by a {\em Fourier} 
transformation
\bea
\label{Fourier}
c_j (t) =  \frac{1}{\sqrt{N}} \sum_k  
\tilde{c}_k (t)\, e^{-i k (ja)}
\, ,
\ea
followed by a {\em Bogoliubov} transformation \cite{bogoliubov}. 
Here $a$ is lattice spacing.
The Bogoliubov transformation
\bea
\label{Bogoliubov-0}
\tilde c_k (t) 
= 
u_k (t) \,  \, \gamma_k 
+ iv_k ^* (t) \, \, \gamma _{-k} ^{\dag} 
\ea
maps the Hamiltonian into a new set of fermionic operators $\gamma_k$ 
whose number is conserved. 
The same anti-commutation relations as in Eqn.~(\ref{jordan-wigner2})
are also satisfied by $\gamma_k$ and $\gamma _k ^{\dag} $
\bea
\label{jordan-wigner3}
\left\{\gamma_k, \gamma_{k'}^{\dag} \right\}=\delta_{kk'}
\quad
,
\quad
\left\{\gamma_k, \gamma_{k'} \right\}
=
\left\{ \gamma_k^{\dag} ,\gamma_{k'}^{\dag} \right\}
=0
\, .
\ea
Since these fermionic operators are supposed to be time-independent, the Bogoliubov 
coefficients $u_k$ and $v_k$ must satisfy \cite{dziarmaga} the equations of motion 
\bea
\label{Motion}
&&
i\frac{du_k}{dt} 
=  
\alpha_k (t) u_k (t) \, 
+ \,  \beta_k (t) v_k (t) \,, 
\nn
&&
i\frac{dv_k}{dt} 
=  
- \alpha_k (t) v_k (t) \, 
+ \,  \beta_k (t) u_k (t) \,,  
\ea
where 
%
$
\alpha_k=2 - 4 g(t)  \cos^2 \left(k a/2\right)
\,
,
\beta_k=  2 g(t) \sin(k a)$.
%
For an adiabatic evolution 
\mbox{$\bra{\psi_s}\dot{H}_{\rm{sys}}\ket{\psi_r}\ll \left(\Delta E_{sr}\right)^2$},
these equations of motion can be solved approximately
\bea
\label{Bogoliubov}
u_k (t) 
&\approx& 
\frac{\alpha_k (t) + \mathscr{E}_k (t)}{{\cal N}_k} 
\exp\left\{ -i \int\limits_0^t \, dt'\mathscr{E}_k (t') \right\}\,,\nn
v_k (t)  
&\approx& 
 \,
\frac{\beta_k (t)}{{\cal N}_k}
\,
\exp\left\{ -i \int\limits_0^t \, dt'\mathscr{E}_k (t') \right\}
\ea
with the normalization 
\mbox{${\cal N}_k=\sqrt{2\mathscr{E}_k^2 + 2\alpha_k\mathscr{E}_k}$} 
ensuring 
\mbox{$|u_k|^2 + |v_k|^2 = 1$} 
and the single-particle energies 
\bea
\label{single-particle}
\mathscr{E}_k(t) = 2\sqrt{1-4g(t)\left[1-g(t)\right]\cos^2\left(ka/2\right)} 
\,.
\ea
All the excitation energies $\mathscr{E}_k$ take their minimum values
%
$\mathscr{E}_k^{\rm min}=2\left|\sin\left(ka/2\right)\right|$,
%
at the critical point 
\mbox{$g_{\rm cr}=1/2$}.   
The pseudo-momenta $ka$ take half-integer values
\mbox{$ka\in(1+2{\mathbb Z})\pi/N\;:\;|ka|<\pi$}.
In view of the \mbox{$k$-spectrum} the minimal gap 
between the ground state and the first excited state scales as 
\mbox{$\Delta E_{\rm min}=\ord(1/N)$}. 
Finally, the Hamiltonian~(\ref{Hamiltonian-Ising}) 
in the subspace of an even number of quasi-particles reads 
\bea
\label{HFinal}
{H}_{\rm{sys}}(t) = \sum_k \mathscr{E}_k(t) 
\left(\gamma_k^{\dag}\gamma_k -\frac{1}{2} \right)
\ea
with fermionic creation and annihilation operators 
$\gamma_k^{\dagger} , \gamma_k$.
Hence, its (instantaneous) ground state contains no fermionic
quasi-particles 
\mbox{$\forall_k\;:\;{\gamma}_k\ket{\psi_0(t)}=0$}. 
Without the environment, the number of fermionic quasi-particles 
\mbox{$\gamma_k^{\dag}\gamma_k$} would be conserved and the system would stay
in an eigenstate (e.g., ground state) for an adiabatic evolution. 
The coupling to the bath, however, may cause excitations and thus the
creation of quasi-particles due to decoherence. 
%


Of course, the impact of decoherence depends on the properties of the
bath and its interaction with the system (decoherence channels).
In the following, we study three different decoherence channels.
However, in all of these different cases, we do not specify the bath
${H}_{\rm{bath}}$ in much detail for the purpose of deriving generally
applicable results.

\subsection{Uniform Coupling Strengths} 

Let us first consider an interaction
\mbox{$\lambda \, {H}_{\rm{int}}$} which is always present:  
In the Hamiltonian ${H}_{\rm{sys}}$ in Eqn.~(\ref{Hamiltonian-Ising}), the
transverse field \mbox{$B(t)=1-g(t)$} appears as a classical control
parameter $B_{\rm cl}$. 
However, the external field \mbox{$B \to B_{\rm cl}+\delta B$} does also
possess (quantum) fluctuations $\delta B$, which couple to the system
of Ising spins.  
Therefore, we start with the following interaction Hamiltonian 
\bea
\label{Interaction1}
H_{\rm{int}} = \left(\sum_j \sigma_{x}^j\right) \otimes \delta B
\,,
\ea
where \mbox{$\delta B$} denotes the bath operator. 
Note that this perturbation should be considered as mild, since it does
not even destroy the bitflip symmetry of the Ising model~(\ref{Hamiltonian-Ising}) and thus does not lead
to leakage between the two subspaces of even and odd bitflip symmetry (or quasi-particle
number, respectively).
This interaction Hamiltonian yields the same matrix
elements as the non-adiabatic corrections 
\mbox{$\bra{\psi_s }\dot{H}_{\rm{sys}}\ket{\psi_r}$} 
in Eqn.~(\ref{Diff}), which can therefore be calculated analogously. 
Insertion of \mbox{$\lambda H_{\rm{int}}$} into Eqn.~(\ref{GAmplitude}) 
yields
\bea          
\label{Amplitude1}          
\mathfrak{A}_s 
&\approx& 
-i\lambda \int d\omega\,
f_s(\omega) 
\int_0^T dt\,
\bra{\psi_s(t)} \sum_j\sigma_j ^x \ket{\psi_0(t)}\times
\nn
&&
\times
\exp\left\{i\left[-\omega t+\int_0^t dt'\,\Delta E_{s0}(t')\right]\right\}    
\,.
\ea       
We may also here include all relevant properties of the environment into
the 
single-operator (compare with Eqn.~(\ref{grover-fourier}) for the double operator version)
spectral function $f(\omega)$ of the bath 
\bea
\label{Fourier1}
e^{-i\Delta\varphi'_s(t)} 
\bra{\alpha_s(t)}\delta B(t)\ket{\alpha_0(t)}  
\equiv           
\int_{-\infty}^{+\infty} d\omega\,e^{-i\omega t}f_s(\omega) 
\,,
\nn
\ea
where $\Delta\varphi'_s$ coincides with $\varphi_s-\varphi_0$ in 
Eqn.~(\ref{Phase}) apart from the system's energy gap $\Delta E_{s0}$
and is typically dominated by the contribution from
\mbox{$H_{\rm{bath}}^{ss}-H_{\rm{bath}}^{00}$}.  
Note that  Eqn. (\ref{Fourier1}) is the
generalization of Eqn. (\ref{grover-fourier}) for the case that the bath state changes strongly.
As a first approximation, we assume that $f(\omega)$ does not change
significantly if we increase the system size $N$ (scaling limit).
After inserting the Jordan-Wigner,
Fourier, and Bogoliubov-transformations, the
matrix element in Eqn.~(\ref{Amplitude1}) reads 
%
%
\begin{figure}
\includegraphics[height=6cm,clip=true]{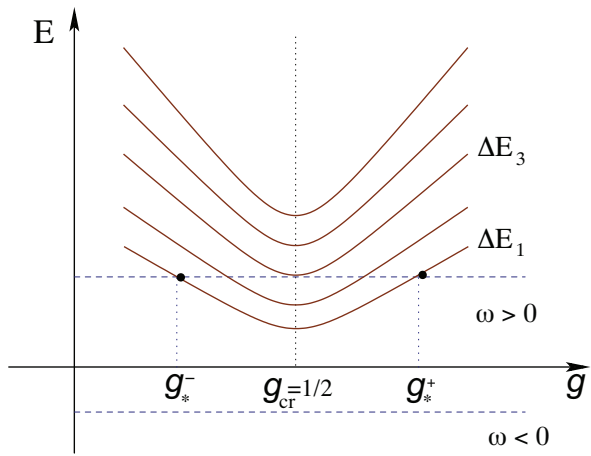}
\caption
{\label{Energy0}
(Color Online)
 Sketch of the excitation spectrum of the Ising 
  chain ${H}_{\rm{sys}}$ as a function of $g$. For a given frequency
  $\omega>0$, real saddle points correspond to intersections of the
  (solid) energy level curves (e.g., $\Delta E_1$) with the (dashed)
  vertical $\omega$-line which occur shortly before ($g_*^-$) and 
  after ($g_*^+$) the quantum phase transition at $g_{\rm
  cr}=1/2$. The saddle-point approximation can   
  only be applied if the intersection angle is large enough, i.e., for
  the drawn $\omega>0$ line, it would work for $\Delta E_1$, but not
  for $\Delta E_3$ etc.   
} 
\end{figure} 
\bea
\label{Element}
\sum_j\bra{\psi_s}\sigma_x^j(t)\ket{\psi_0} 
\approx
\frac{2ig(t)\sin(ka)}{\mathscr{E}_k(t)}
\bra{\psi_s}{\gamma}_k^{\dag}{\gamma}_{-k}^{\dag}\ket{\psi_0} 
\,,
\nn
\ea
where the $\approx$ sign refers to the adiabatic approximation. 
Thus, it is only non-vanishing for excited states $\ket{\psi_s}$
containing two quasi-particles $s=(k,-k)$ with opposite momenta and
hence we get $\Delta E_{s0}=2\mathscr{E}_k$. 
%


First of all, in order to have a {\em quantum} phase transition 
(or a working adiabatic quantum computer), the environment should be
cold enough to permit the preparation of the system in the initial
ground state such that $f(\omega)$ is only non-negligible when
\bea
\label{w0}
\omega\ll2=\mathscr{E}_k(t=0)
\ea
holds, compare also Eqn.~(\ref{Efourierexample}).
Therefore, we will analyze the spectral excitation amplitude 
\mbox{$\mathfrak{A}_s^\omega$}
defined via 
\bea
\label{spectral-Amplitude0}
\mathfrak{A}_s\equiv\int d\omega\,f_s(\omega)\,
\mathfrak{A}_s^\omega
\ea
in the different $\omega$-regimes in the following.

%
\subsubsection{Intermediate Positive Frequencies}
We may solve the time integral via the saddle-point (or stationary phase) 
approximation for intermediate positive frequencies,
\bea
\label{w1}
2\gg\omega\gg\Delta E_{s0}^{\rm min}\approx2|ka|
\, .
\ea
For the exponent in Eqn.~(\ref{Amplitude1}) the saddle-point condition reads
\bea
\label{SCondition}
\left[ \pdiff{h_k(t, \omega)}{t}\right]_{t=t_*} 
\hspace{-0.3cm}
= 0 
\, \, \leadsto \, \, 
\omega=\Delta E_{s0}(t_*)=2\mathscr{E}_k(t_*) \, , 
\ea
where  
\mbox{$ h_k(t, \omega) = i \left[-\omega t
+2\int_0 ^t dt'\,\mathscr{E}_k(t')\right] $}
and $t_*$ denotes the saddle points.
This condition yields two saddle points shortly before and after the
transition, see also Fig.~(\ref{Energy0})
\bea
\label{SPoints}
g(t_*^\pm)
=
\frac{1}{2} 
\pm 
\frac{\left[\omega ^2 - 16 \sin^2 
\left(ka/2\right)\right]^{1/2}}
{8 \cos\left(ka/2\right)} 
\, . 
\ea
The saddle-point approximation yields for the spectral excitation amplitude defined in~(\ref{spectral-Amplitude0})
\bea
\label{spectral-Amplitude}    
\mathfrak{A}_s^{\omega\gg2|ka|}
&\approx&
\sqrt{\frac{\pm32\pi i\lambda^2  \sin(ka)\sin\left(\frac{ka}{2}\right)e^{2h_k(t_*, \omega)}}
{\omega \dot{g} (t_*)  g^{-2}(t_*) 
\sqrt{\omega^2 - 16 \sin^2 
\left(\frac{ka}{2}\right)
}}}\nn
&&
+\ord\left(\frac{\lambda\dot g(t_*)}
{\omega\,\sqrt{\omega^2-4k^2a^2}}\right)
\, ,
\ea
which depends on the interpolation dynamics $g(t)$. 
The minimum gap can be obtained from Eqn.~(\ref{single-particle})
and does indeed scale polynomially \mbox{$\Delta E_{\rm min}=\ord(1/N)$}
and, thus:
\begin{enumerate}
\item[-]
For a constant speed interpolation \mbox{$g(t)=t/T$},
the necessary run-time for an adiabatic evolution  $T$ 
scales polynomially \mbox{$T=\ord\left(\Delta E^{-2}_{\rm min}\right)
=\ord(N^2)$}.
\item[-]
For adapted interpolation dynamics 
\mbox{$\dot g(t)\propto\Delta E(t)$}
or 
\mbox{$\dot g(t)\propto\Delta E^2(t)$,}
however, one may achieve shorter run-times of 
\mbox{$T=\ord(N\ln N)$}
or 
\mbox{$T=\ord(N)$,}
respectively \cite{schaller2006b}
and therefore better results for the spectral excitation amplitude,
see \mbox{Table~\ref{Table}}.
\end{enumerate}

\subsubsection{Near the Minimum Gap}

From Eqn.~(\ref{spectral-Amplitude}) it follows that the
saddle-point approximation breaks down if 
$\omega$ approaches the minimum gap $w\approx\Delta E_{s0}^{\rm min}\approx2|ka|$, 
see Fig.~(\ref{Energy0}). 
In this case, we may obtain an upper bound for the time integral in 
Eqn.~(\ref{Amplitude1}) via omitting all phases.
For a constant speed interpolation \mbox{$g(t)=t/T$}
\bea
\label{upper-bound}    
\mathfrak{A}_s^{\omega\approx2|ka|}
&\le&
\frac{2\lambda\sin(ka)}{T}
\int_0^T dt \, \frac{t}{\mathscr{E}_k(t)}\nn
&=&
\ord\left(\lambda N^2\omega\ln \omega\right)
\,.
\ea
Similarly, one can get better results for 
adapted interpolation dynamics, see Table~\ref{Table}. 
%
\begin{table}[t]
\begin{center}
\begin{tabular}{|c|c|c|}
\hline
&
$1\gg\omega\gg2ka$
&
$1\gg\omega\approx2ka$
\\
\hline
$\ddot g(t)=0$
&
$\ord\left(\lambda ka\omega^{-1}N\right)$ 
&
$\ord\left(\lambda N^2\omega\ln \omega\right)$ 
\\
$\dot g(t)\propto\Delta E(t)$
&
$\ord\left(\lambda ka\omega^{-3/2}\sqrt{N}\right)$
&
$\ord\left(\lambda N\ln N \right)$
\\
$\dot g(t)\propto\Delta E^2(t)$
&
$\ord\left(\lambda ka\omega^{-2}\right)$
&
$\ord\left(\lambda N\right)$ 
\\
\hline
\end{tabular}
\end{center}
\caption{\label{Table} {\small{Scaling of the spectral excitation amplitude
  \mbox{$\mathfrak{A}_s^\omega$} in the saddle-point approximation
  (\mbox{$\omega\gg2ka$}) and its upper bound (\mbox{$\omega\approx2ka$}) for
  different interpolation dynamics $g(t)$, where \mbox{$\Delta
  E(t)=2\mathscr{E}_{k=\pi/(aN)}(t)$} denotes the fundamental gap. In all		
  cases, the total excitation probability (integral over all $\omega$ and sum over all 
  $k$) increases with system size $N$}}.}
\end{table}

\subsubsection{Positive Frequencies Below the Minimum Gap}

For positive frequencies fulfilling 
\mbox{$0 \le \omega \ll 2|ka|$}, the saddle points at  
\bea
\label{w3}
g(t_*^\pm)\approx
\frac{1}{2}\, \pm\, \frac{1}{8}
\sqrt{\omega^2-4k^2a^2}
\, ,
\ea
move away from the real axis and thus the exponent in 
Eqn.~(\ref{Amplitude1}) contains real terms. The constant speed
interpolation leads to 
\bea
\label{w4}
i \left[-\omega t_*+2\int_0 ^{t_*} dt\,\mathscr{E}_k(t)\right]
\approx i\eta-
\left[\frac{\omega|ka|}{4}+\frac{(ka)^2}{2}\right]T
\, ,
\nn
\ea
where $\eta$ is a real value. Therefore, 
the spectral excitation amplitude is exponentially suppressed in the
adiabatic limit
\bea
\label{suppressed} 
\mathfrak{A}_s^{\omega\ll2|ka|}
=\ord\left(\exp\left\{-\frac12\,T(ka)^2\right\}\right)
\,.
\ea
%

%
%
\subsubsection{Negative Frequencies}
\begin{figure}
\includegraphics[height=6cm,clip=true]{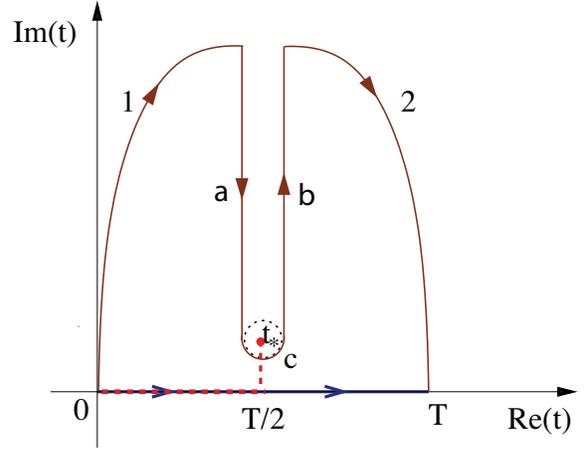}
\caption
{\label{Contour} 
(Color Online)
A sketch of the deformed integration contour.
The original integration contour
(blue line along the real axis) is shifted to the complex plane 
(curved line), where
$t_*$ indicates the singular point,  
$t_* = T/2 + i T/2 \tan\left(ka/2\right)$.
Only paths $a$ and $b$ contribute significantly to the integral.
}
\end{figure}

Finally, for negative frequencies $\omega<0$, the saddle points
collide with the branch cut generated by the square-root in  
$\mathscr{E}_k$.
In this case, we may also estimate the spectral excitation amplitude 
\mbox{$\mathfrak{A}_s^\omega$}
in Eqn.~(\ref{spectral-Amplitude0})
by deforming the time integration contour into the complex plane.
We assume that all involved functions can be analytically
continued into the complex plane and are well-behaved
near the real axis. 
Given this assumption, we deform the integration contour into the 
upper complex half-plane to obtain a negative exponent which is the
usual procedure in such estimates until reaching a saddle point, 
a singularity, or a brunch cut, see Fig.~(\ref{Contour}).
Deforming the integration contour into the lower
complex half-plane would of course not change the result,
but there the integrand is exponentially large and strongly
oscillating such that the integral is hard to estimate.
Since the integral in the complex plane is zero around path $c$
and the integrals on the paths 1 and 2 cancel each other, only 
paths $a$ and $b$ give the main contribution to the integral. 

Let us first consider a constant interpolation function
\mbox{$g(t)=t/T$} which leads to singular points 
%
%
%
%
\bea
\label{singula-point}
t_* = \frac{T}{2}\pm i \frac{T}{2} 
\tan\left(\frac{ka}{2}\right)
\, , 
\ea
in the complex plane.
Performing the time integral in the exponent of 
Eqn.~(\ref{Amplitude1}) acquires a large negative real term in
the exponent
\bea
\label{Exponent}
\int_0^t dt'\,\Delta E_{s0}(t')
&=&
\left\{\int_0^{T/2} 
+\int_{T/2}^{t_*} 
+\int_{t_*}^t \right\} 
dt'\,\Delta E_{s0}(t')
\nn
&\approx&
\xi' \pm 2t^2
+ 
\frac{i\pi T}{16} (ka)^2
\, ,
\ea
where $\xi'$ is a constant and real value. 
Insertion of Eqn.~(\ref{Exponent}) into Eqn.~(\ref{Amplitude1}) 
and doing some algebra yields the exponential suppression for the amplitudes 
in the upper complex half-plane
\bea
\label{Negative-Frequency}
\mathfrak{A}_s^{\omega<0}
&\approx&
\exp
\left\{-\frac{\pi T}{16}\left(ka\right)^2
\right\} \, \times
\nn
&&
\times \,
\int_{a,b} dt \, F(t)
\exp\Big\{i\left(2t^2-\omega t+\tau \right)\Big\}
\ea
with \mbox{$F(t) =\bra{\psi_s(t)} \sum_j\sigma_j ^x \ket{\psi_0(t)}$}
and where $\tau$ is a real constant.
Therefore, applying the inequality 
\bea
\label{inequality00}
|\mathfrak{H}|\le\int dy \, |\phi (y)|
\quad
{\rm with}
\quad
\mathfrak{H}=\int dy\,  \phi (y)
\, ,
\ea
the amplitudes for negative frequencies are also exponentially suppressed
for $g(t)=t/T$ and similarly for the other interpolations. 
This result can be understood in the following way:
For frequencies $\omega$ below the lowest excitation energies, the
energy $\omega$ of the reservoir modes is not sufficient for exciting 
the system via energy-conserving transitions. 
Hence excitations can only occur via non-adiabatic processes 
for which energy-conservation becomes ill-defined, but these
processes are suppressed if the evolution is slow enough. 

An estimate on the total error probability $|\mathfrak{A}_s|^2$ introduced in 
Eqn.~(\ref{spectral-Amplitude0}) is obtained by 
performing the weighted sum of the contributions from the different $\omega$-regimes,
which depends on the Fourier transform of the bath correlation function
\bea
\label{total-error}
\abs{\mathfrak{A}_s^{\omega}} &<& 
\left[\max\limits_{\omega\gg 2|ka|} \abs{\mathfrak{A}_s^{\omega \gg 2|ka|}}\right] \int\limits_{2 \gg \omega \gg 2 |ka|} \abs{f_s(\omega)} d\omega\nn
&&
+\left[\max\limits_{\omega\approx 2|ka|} \abs{\mathfrak{A}_s^{\omega\approx  2|ka|}}\right] \int\limits_{\omega \approx |ka|} \abs{f_s(\omega)} d\omega\nn
&&+\left[\max\limits_{0 < \omega< 2|ka|} \abs{\mathfrak{A}_s^{\omega \ll 2|ka|}}\right] \int\limits_0^{2|ka|} \abs{f_s(\omega)} d\omega\nn
&&+\left[\max\limits_{\omega<0} \abs{\mathfrak{A}_s^{\omega<0}}\right] \int\limits_{-\infty}^{0} \abs{f_s(\omega)} d\omega\,.
\ea
From the results in Eqns.~(\ref{spectral-Amplitude}), (\ref{upper-bound}), (\ref{suppressed}), and (\ref{Negative-Frequency})
it becomes obvious that although the last two contributions in the above sum can be efficiently suppressed, the
first two terms will scale with the system size.
Therefore, for the adiabatic Ising model, decoherence can
only be effectively suppressed when the bath spectral function $f_s(\omega)$ has
only support at frequencies below the minimum fundamental energy gap.
For a bosonic bath in thermal equilibrium this would imply a reservoir
temperature below the minimum fundamental energy gap, compare also 
Eqn.~(\ref{Efourierexample}).


\subsection{Nonuniform Coupling Strengths} 

In a realistic situation, the Ising spins will not be symmetrically coupled to the environment
\bea
\label{Interaction2}
H_{\rm{int}} = \sum_j \sigma_{x}^j \otimes \delta B_j
\,,
\ea
where \mbox{$\delta B_j$} denote now different operators acting on the bath Hilbert space.
Insertion of~(\ref{Interaction2}) into Eqn.~(\ref{GAmplitude})
and evaluating the corresponding matrix element by applying the Jordan-Wigner,
Fourier, and Bogoliubov transformations of Pauli matrices -- given in 
Eqns.~(\ref{Jordan}),~(\ref{Fourier}), and~(\ref{Bogoliubov-0}) -- yield
\bea          
\label{Amplitude2}          
\mathfrak{A}_s 
&\approx& 
-\frac{2i\lambda}{N} \sum_{k,k'} \int d\omega\,
f_{k,k'}^{s}(\omega) 
\int_0^T dt\,\, u_{k}^*(t) \, v_{k'}(t)
\times
\nn
&&
\hspace{-1.2cm}
\bra{\psi_s(t)} \gamma_{k}^{\dagger}\,\gamma_{-k'}^{\dagger} \ket{\psi_0(t)}
\,
\exp\left\{i\left[-\omega t+\int_0^t dt'\,\Delta E_{s0}(t')\right]\right\}    
\,,
\nn
\ea       
where we include again all relevant properties of the environment into
the spectral function $f_{k,k'}^s(\omega)$ of the bath 
\bea
\label{Fourier2}
&& 
\hspace{-0.5cm}
e^{-i\Delta\varphi'_s(t)} \left(\sum_j \delta B^{s0}_j
\,e^{i(k-k')ja} \right)
\equiv        
\int_{-\infty}^{+\infty} 
\hspace{-0.2cm}
d\omega\,e^{-i\omega t}f_{k,k'}^s(\omega) 
\,,
\nn
\ea
where $\delta B^{sr}_j =\bra{\alpha_s(t)}\delta B_j(t)\ket{\alpha_r(t)}$,
$\Delta\varphi'_s$ coincides with $\varphi_s-\varphi_0$ in 
Eqn.~(\ref{Phase}) apart from the system's energy gap $\Delta E_{s0}$
and $u_k(t)$, $v_k(t)$ are the Bogoliubov coefficients given in 
Eqn.~(\ref{Bogoliubov}).
The excitation amplitude in Eqn.~(\ref{Amplitude2}) is only non-vanishing
for the excited states containing two quasi-particles $s=(k,-k')$
\bea
\label{GECH2}
\Delta E_{s0}= \mathscr{E}_k+\mathscr{E}_{k'}
\, .
\ea
Insertion of $\displaystyle{u_{k}^*}$, $\displaystyle{v_{k'}}$ given
by Eqn.~(\ref{Bogoliubov}), into Eqn.~(\ref{Amplitude2}) yields
\bea          
\label{Amplitude3}          
\mathfrak{A}_s 
&\approx& 
\frac{i\lambda}{N}\sum_{k,k'} \int d\omega\,
f_{k,k'}^{s}(\omega) 
\int_0^T dt
\bra{\psi_s(t)} \gamma_{k}^{\dagger}\,\gamma_{-k'}^{\dagger} \ket{\psi_0(t)}\times\nn
&&
\times
\,
\frac{{\cal C}_{k,k'} (t)}{ {\cal N}_{k'}}
\,
\exp\left\{i\left[-\omega t+
2\int_0^t dt'\,\mathscr{E}_k(t')\right]\right\}    
\,,
\ea       
where
%
\bea
\label{Coff-Amp}
{\cal C}_{k,k'} (t)=4\,
g(t)\, \sin(k'a)\sqrt{\frac{1}{2}+\frac{1-2g(t)\cos^2(ka/2)}{\mathscr{E}_{k}(t)}} 
\,.
\nn
\ea
Following the same procedure outlined above, we may consider
different domains of $\omega$ in order to solve the time integral in 
Eqn.~(\ref{Amplitude3}).
For the intermediate positive frequencies,
%
\bea
\label{IPF}
2 \gg \omega \gg 
2\,|ka|  \, \, \,  {\rm{and}} \, \, \,  2\,|k'a|
\, ,
\ea
the saddle-point approximation can be applied once again
%
\bea
\label{SP2}
&&
-\omega+ \, 2\, \mathscr{E}_k(t_*)
=0
\, ,
\ea
where $t_*$ denotes the saddle points.
The saddle-point approximation yields for the spectral excitation amplitude
\bea
\label{SA2}
\mathfrak{B}_s^{\omega\gg2|ka|}
=
\ord\left(\frac{\lambda k'a}{N\sqrt{\omega\,\dot g(t_*)\sqrt{\omega^2-4{k}^2a^2}}}\right)
\, ,
\ea
where $\displaystyle{\mathfrak{B}_s^{\omega}}$ is defined as following
\bea
\label{Definition-SA}
\mathfrak{A}_s  \approx \sum_{k,k'}\int d\omega\,
f_{k,k'}^{s}(\omega)\, \mathfrak{B}_s^{\omega}  
\, .
\ea
The spectral excitation amplitude in Eqn.~(\ref{SA2}) depends on the interpolation
dynamics $g(t)$ and is very similar to Eqn.~(\ref{spectral-Amplitude}) apart
from 
$N$ in the denominator.
Presence of $N$ in the denominator may cause 
some slow down for the
error probability, see Table \ref{Table}.
However, existence of many excited states -- 
sum over all possible $k$ and $k'$ in Eqn.~(\ref{Amplitude3}) --
causes the growth of the error probability with the system size.
If $\omega$ approaches the $2|ka|$, the saddle-point approximation breaks down
and we can get a similar upper bound which is shown in Table.~\ref{Table},
by omitting all phases.
With the same argument given above, the amplitudes are exponentially suppressed
in the adiabatic limit for frequencies far below $2|ka|$.
Thus, as with the discussion in the previous subsection below Eqn.~(\ref{total-error}) 
we may conclude that in a more general case where the coupling strength to the bath 
is not 
uniform, 
the impact of decoherence for the environmental
noise is very similar to the coherent reservoir and the error
probability increases with the system size, 
unless the bath temperature lies below the minimum energy gap.

\subsection{Perturbing the Bitflip Symmetry}

Unfortunately, interactions with the reservoir cannot be tailored, such that
one may also expect perturbations that do not reflect the bitflip symmetry of
the Ising Hamiltonian and thereby lead to leakage between the subspaces of 
even and odd quasi-particle numbers.
Let us 
consider a simple case, where only one Ising spin is coupled 
to the environment
\bea
\label{Interaction3}
H_{\rm{int}} = \sigma_{z}^1 \otimes \delta B
\,.
\ea
Insertion of this interaction Hamiltonian into Eqn.~(\ref{GAmplitude}) 
and evaluating the corresponding matrix element yield the excitation amplitude
which is a combination of two terms 
%
\bea
\label{Amplitude3-0}
\mathfrak{A}_s =\, - \, \frac{ i \, \lambda}{\sqrt{N} } \, 
\int d\omega\, f_{s}(\omega) \sum_{k} \left(\mathfrak{A}^{\omega}_{s,1} +\mathfrak{A}^{\omega}_{s,2} \right)
\, ,
\ea
 with
\bea          
\label{Amplitude3-1}          
\mathfrak{A}^{\omega}_{s,1} 
&\approx& 
i\,  e^{-ika}\, \int_0^T dt\, \frac{\beta_k}{{\cal N}_{k}}
\,
\bra{\psi_s(t)} \gamma_{-k}^{\dagger} \ket{\psi_0(t)}
\,e^{-i\omega t}
\nn
\ea 
and
\bea          
\label{Amplitude3-2}          
\mathfrak{A}^{\omega}_{s,2} 
&\approx& 
e^{ika} \,
\int_0^T dt\,
\sqrt{\frac{1}{2}+\frac{1-2g(t)\cos^2(ka/2)}{\mathscr{E}_k(t)}}
\times
\nn
&&
\hspace{-0.9cm}
\times \, \bra{\psi_s(t)} \gamma_{k}^{\dagger} \ket{\psi_0(t)} \exp\left\{i\left[-\omega t+2\int_0^t dt'\,
\mathscr{E}_k(t')\right]\right\}  
\,,
\nn
\ea
where the spectral function $f_s(\omega)$ of the bath is as defined in Eqn.~(\ref{Fourier1}).
%
%
For large $T$, the first term $\mathfrak{A}^{\omega}_{s,1} $ is a Fourier transformation of some
function of $\omega$  
\bea
\label{Amplitude3-1-1}
\mathfrak{A}^{\omega}_{s,1} 
\approx 
ie^{-ika} \sin(ka) \underbrace{\left(\Sigma(\omega)
-\underbrace{\int_{-T}^0 \hspace{-0.2cm} 
dt \,\, \Xi(t)\, e^{-i\omega t}}_{\ord\left(1/\omega\right)}
\right)}_{\Omega(\omega)}\, ,
\nn
\ea
where
\bea
\label{FF1}
\Xi(t)=\frac{2g(t)}{\sqrt{2 \mathscr{E}^2_k(t)+
4\left[1-2g(t)\cos^2(ka/2)\right]\mathscr{E}_k(t)}},
\ea
%
%
$\Sigma(\omega)=\int_{-T}^T  dt \, \Xi(t)\, e^{-i\omega t}$,
and $\Omega(\omega)$ is some function of $\omega$.
In order to solve the time integral of $\mathfrak{A}^{\omega}_{s,2} $,
we may employ saddle point approximation which yields 
\bea
\label{Amplitude3-2-1}
\mathfrak{A}^{\omega}_{s,2}= \ord(T)\, .
\ea
Thus, we can conclude that
\bea
\label{Amplitude3-3}
\mathfrak{A}_s =\int d\omega\,
f_{s}(\omega) \sum_{k} \underbrace{\left[\ord\left(\frac{\lambda ka}{\sqrt{N}}\right)
+\ord\left(\frac{\lambda T}{\sqrt{N}}\right)\right]}_{\mathfrak{A}^{\omega}_s} \,.
\ea
The optimal run-time for an adiabatic evolution $T$ depends on the 
interpolation dynamics $g(t)$, see \cite{schaller2006b}.
Scaling of the spectral excitation amplitude $\mathfrak{A}^{\omega}_s$ for different 
interpolation dynamics is shown in Table~\ref{Perturb-Bitflip}.
%
\begin{table}[t]
\begin{center}
\begin{tabular}{|c|c|}
\hline
&
$\mathfrak{A}^{\omega}_s$
\\
\hline
$\ddot g(t)=0$
&
$\ord\left(\lambda ka/\sqrt{N}\right)+
\ord\left(\lambda N^{3/2}\right)$
\\
$\dot g(t)\propto\Delta E(t)$
&
$\ord\left(\lambda ka/\sqrt{N}\right)+
\ord\left(\lambda\sqrt{N} \ln N\right)$
\\
$\dot g(t)\propto\Delta E^2(t)$
&
$\ord\left(\lambda ka/\sqrt{N}\right)+
\ord\left(\lambda \sqrt{N}\right)$
\\
\hline
\end{tabular}
\end{center}
\caption{\label{Perturb-Bitflip} {\small{Scaling of the spectral excitation amplitude
  \mbox{$\mathfrak{A}_s^\omega$} for
  different interpolation dynamics $g(t)$, where \mbox{$\Delta
  E(t)=\mathscr{E}_{k=\pi/(aN)}(t)$} denotes the fundamental gap. In all		
  cases, the total excitation probability (integral over all $\omega$ and sum over all 
  $k$) increases with system size $N$}}.}
\end{table}
This decoherence channel poses a significant problem to robust
ground state preparation in the Ising model:
Regardless how low the bath temperature is, the decoherent excitation probability 
will scale with the system size!
This finding is consistent with the experience that large Schr\"odinger cat states
as~(\ref{Esupercat}) are extremely sensitive to decoherence.

\section{Summary}
%

In summary, we studied the 
impact of decoherence due to a weak coupling to a rather general environment on
first and second order quantum phase transitions.
%
Since the Ising model is considered \cite{sachdev} as a prototypical
example for a second-order quantum phase transition, we expect our
results to reflect general features of second-order transitions.    

For the decoherence channel~(\ref{Interaction1}) which is always
present (though possibly not the dominant channel), we already found
that the total excitation probability 
always increases with system size $N$ (continuum limit): 
Even though the probability for the {\em lowest} excitation
$k=\pm\pi/(aN)$ can be kept under control for a bath which is
well-behaved in the infra-red limit -- see also Sec.~(3.4) -- the
existence of {\em many} excited states 
$k\in\pi(1+2{\mathbb Z})/(aN)\;:\;|ka|<\pi$
converging near the critical point causes the growth of the error
probability for large systems.  
This growth can be slowed down a bit via adapted interpolation schemes
$g(t)$, but not stopped. 

Other decoherence channels will 
in the best case 
display the same general behavior:
E.g., for $2 \gg \omega\gg|ka|$, the associated amplitudes scale as 
\bea
\label{amplitude01}
\mathfrak{A}_s^\omega=\ord\left(\frac{\lambda\phi_s(t_*)}
{\sqrt{\dot g(t_*)}}\right)
\, ,
\ea
where $\phi_s$ denotes the matrix element in analogy to~(\ref{Element}). 
Typically, for a homogeneous coupling to the bath, $\phi_s$ does not
strongly depend on the system size $N$ (for given $ka$ and $\omega$). 
Since $\dot g(t_*)$ decreases for $N\to\infty$ or at least
remains constant -- for $\dot g(t)\propto\Delta E^2(t)$ -- the total
excitation probability again increases with system 
size $N$.
%
If only a few spins are coupled via $\sigma^x_j$ to the environment, the
matrix element  $\phi_s$ in Eqn.~(\ref{Amplitude2}) 
will 
decrease  with the system size $\ord(1/N)$ and then the error probability may be kept
under control -- for $\dot g(t)\propto\Delta E^2(t)$.
However, $\phi_s$  is of order $\ord(1/\sqrt N)$ for the the $\sigma^z$-channel 
given in Eqn.~(\ref{Amplitude3-0}) and 
the Schr\"odinger cat states are still sensitive to decoherence.

According to the analogy between adiabatic quantum algorithms and quantum
phase transitions \cite{latorre,gernot}, this result suggests
scalability problems of the corresponding adiabatic quantum algorithm
-- unless the temperature of the bath stays below the ($N$-dependent)
minimum gap \cite{childs_robust} or the coupling to the bath decreases
with increasing $N$. 
These problems are caused by the accumulation of many levels at the
critical point \mbox{$g_{\rm cr}=1/2$}, 
which presents the main difference to isolated avoided level crossings
(corresponding to first-order phase transitions) discussed earlier. 
It also causes some difficulties for the idea of thermally assisted
quantum computation (see, e.g., \cite{amin}) since, in the presence of
too many available levels, the probability of hitting the ground
state becomes small.

Therefore, in order to construct a scalable adiabatic quantum
algorithm in analogy to the Ising model, suitable error-correction
methods will be required. 
As one possibility, one might exploit the quantum Zeno effect 
and suppress transitions in the system by constantly measuring the
energy, see for example \cite{childs_zeno}. 
Another interesting idea are adiabatic ground state preparation schemes 
(algorithms) that provide a constant lower bound on the fundamental energy gap 
that does not scale with the system size.
In case of the Ising model discussed here, this is possible by increasing
the complexity of the interpolation path (i.e., beyond
the straight-line interpolation).
Unfortunately, the simplest approach to the Ising model \cite{gernot-nonlinear} only bounds 
the fundamental gap in the subspace of even bitflip parity, 
i.e., decoherence channels that mediate transitions between the two subspaces as 
e.g. in Eqn.~ (\ref{Interaction3}) will destroy the associated robustness against decoherence.
Many-body interactions in the system Hamiltonian are required to resolve this problem.
%
%
%
In this case, decoherence could be strongly suppressed for a
low-temperature bath.  
Of course, the generalization of all these concepts and results to
more interesting cases such as the (NP-complete) two-dimensional Ising
model is highly non-trivial and requires further investigations.

%
\section{Acknowledgments}
%
This work was supported by the DFG
(SCHU~1557/1-2,3; SCHU~1557/2-1; SFB-TR12).

$^*$\,{\footnotesize\sf ralf.schuetzhold@uni-due.de}\,\,   


\begin{thebibliography}{9999}
\newfont{\cyr}{wncyr10}

\bibitem{sachdev} 
S.~Sachdev, 
{\em Quantum phase transitions},
(Cambridge University Press, Cambridge, UK, 1999).

\bibitem{nielsen}
M.~A. Nielsen and I.~L. Chuang,
{\em Quantum computation and quantum information},
(Cambridge University Press, Cambridge,  England, 2000).

\bibitem{farhi-2000}
E.~Farhi, J.~Goldstone, S.~Gutmann and M.~Sipser,
pre-print: {\tt quant-ph/0001106} (2000).

\bibitem{childs_robust} 
A.~M.~Childs, E.~Farhi, and J.~Preskill,
Phys.~Rev.~A {\bf 65}, 012322 (2001).

\bibitem{sarandy_open} 
M.~S.~Sarandy, and D.~A.~Lidar,
Phys.~Rev.~Lett.~{\bf 95}, 250503 (2005).


\bibitem{messiah}
A.~Messiah,
{\em Quantum mechanics},
(John wiley and Sons, 1958).

\bibitem{latorre} 
J.~I.~Latorre and R.~Or\'us,
 Phys.~Rev.~A~{\bf 69}, 062302 (2004).

\bibitem{gernot} 
R.~Sch\"utzhold and G.~Schaller,
Phys.~Rev.~A {\bf 74}, 060304(R) (2006).

\bibitem{roland}
J.~Roland and N.~J. Cerf,
Phys.~Rev.~A {\bf 65}, 042308 (2002).

\bibitem{bunder} 
J.~E.~Bunder and R.~H.~McKenzie,
Phys.~Rev.~B {\bf 60}, 344 (1999).


\bibitem{fischer} 
K.~H.~Fischer, and J.~A.~Hertz,
{\em Spin glasses},
(Cambridge University Press, Cambridge, UK,  1993).

\bibitem{chakrabarti} 
A.~Das, and B.~K.~Chakrabarti, 
(LNP~{\bf 679}, Springer-Verlag, Heidelberg, 2005).

\bibitem{santoro} 
G.~E.~Santoro, R.~ Marto\v n\' ak, E.~Tosatti, and R,~Car,  
Science~{\bf 295}, 2427 (2002).

\bibitem{kadowaki}
T.~Kadowaki, and H.~Nishimori,
Phys.~Rev.~E {\bf 58}, 5355 (1998).

\bibitem{dwave} 
{\tt http://tinyurl.com/yoz77v}, see also
W. van Dam, 
Nature Physics {\bf 3}, 220 (2007).

\bibitem{deco} 
S.~Mostame, G.~Schaller, and R.~Sch\"utzhold,
Phys.~Rev.~A {\bf 76}, 030304(R) (2007).

\bibitem{dziarmaga} 
J.~Dziarmaga,
Phys.~Rev.~Lett.~{\bf 95}, 245701 (2005).

\bibitem{gernot-symmetry}
G.~Schaller, and R.~Sch\"utzhold,
Quantum Information and Computation~{\bf 10}, 0109 (2010).

\bibitem{schuetzhold2008a}
R. Sch\"utzhold,
Journal of Low Temperature Physics {\bf 153}, 228-243 (2008).

\bibitem{schaller2008a}
G. Schaller and T. Brandes,
Phys. Rev. A {\bf 78}, 022106 (2008).

\bibitem{brandes2005a}
T. Brandes,
Physics Reports {\bf 408}, 315-474 (2005).

\bibitem{farhi0512159}
E.~Farhi, J.~Goldstone, Sam Gutmann and Daniel Nagaj,
International Journal of Quantum Information {\bf 6}, 503 (2008).

\bibitem{zindaric-NP}
M.~\u{Z}nidari\u{c}, and M.~Horvat,
Phys.~Rev.~A {\bf 73}, 022329 (2006).

\bibitem{group_phase0}
R.~Sch\"utzhold, M.~Uhlmann, Y.~Xu and Uwe R.~Fischer,
Phys.~Rev.~Lett.~{\bf 97}, 200601 (2006).

\bibitem{group_phase1}
R.~Sch\"utzhold,
Phys.~Rev.~Lett.~{\bf 95}, 135703 (2005).

\bibitem{vidal}
G.~Vidal, J.~I.~Latorre, E.~Rico, and A.~Kitaev, 
Phys.~Rev.~Lett.~{\bf 90}, 227902 (2003).

\bibitem{zurek}
W.~H.~Zurek, U.~Dorner, and P.~Zoller, 
Phys.~Rev.~Lett.~{\bf 95}, 105701 (2005).

\bibitem{damski}
B.~Damski, 
Phys.~Rev.~Lett.~{\bf 95}, 035701 (2005).

\bibitem{sengupta}
K.~Sengupta, S.~Powell, and S.~Sachdev, 
Phys.~Rev.~A~{\bf 69}, 053616 (2004).

\bibitem{fazio}
D.~Patane, L.~Amico, A.~Silva, R.~Fazio, and G.~E.~Santoro, 
Phys.~Rev.~B~{\bf 80}, 024302 (2009).

\bibitem{fubini}
A.~Fubini, G.~Falci and A.~Osterloh,
New~J.~Phys.~{\bf 9}, 134 (2007).

\bibitem{sarandy2004}
M.~S.~Sarandy, L.~A.~Wu, and D.~A.~Lidar,
Quant.\ Inf.\ Proc.\ {\bf 3}, 331 (2004).


\bibitem{markus} 
M.~Tiersch and R.~Sch\"utzhold,
Phys.~Rev.~A~{\bf 75}, 062313 (2007).

\bibitem{schaller2006b}
G.~Schaller, S.~Mostame, and R.~Sch\"utzhold,
Phys.~Rev.~A {\bf 73}, 062307 (2006).

\bibitem{jansen2007a}
S. Jansen, M. B. Ruskai, and R. Seiler,
J. Math. Phys. {\bf 48}, 102111 (2007).

\bibitem{jordan} 
P.~Jordan and E.~Wigner, 
Z.~Phys.~47, 631 (1928).

\bibitem{bogoliubov} 
S.~Katsura, 
Phys.~Rev.~{\bf 127}, 1508 (1962).


\bibitem{amin} 
M.~H.~S.~Amin, Peter J.~Love, and C.~J.~S.~Truncik,
Phys.~Rev.~Lett.~{\bf 100}, 060503 (2008).

\bibitem{childs_zeno}
A.~M.~Childs, E.~Deotto, E.~Farhi, 
J.~Goldstone, S.~Gutmann and A.~J.~Landahl,
Phys.~Rev.~A {\bf 66}, 032314 (2002).

\bibitem{gernot-nonlinear}
G.~Schaller,
Phys.~Rev.~A  {\bf 78}, 032328 (2008)

\end{thebibliography}
\end{document}